\documentclass[useAMS,usenatbib]{mn2e}

\usepackage{longtable}
\usepackage{pdflscape}
\usepackage{graphicx}
\usepackage{pdflscape}
\usepackage{lscape,graphicx,pifont}
\usepackage{color}

\newcommand{\Msun}{\mbox{\,M$_\odot$}}

\newcommand{\vunit}{\mbox{\,km\,s$^{-1}$}}
\newcommand{\mic}{\mbox{$\,\mu$m}}

\newcommand{\nucl}[2]{\mbox{$^{#1}${#2}}}

\newcommand{\ltsimeq}{\raisebox{-0.6ex}{$\,\stackrel
        {\raisebox{-.2ex}{$\textstyle <$}}{\sim}\,$}}

\newcommand{\tcrb}{\mbox{T~CrB}}

\usepackage{graphicx}	
\usepackage{amsmath}	
\usepackage{amssymb}	

\title[Search for maser emission]{Stringent limits on \nucl{28}{Si}O maser 
   emission from the recurrent nova T~Coronae Borealis}
\author[A. Evans et al.]{A. Evans$^{1}$\thanks{E-mail: a.evans@keele.ac.uk},
B. Pimpanuwat$^{2}$,
 A. M. S. Richards$^{2}$,
D. P. K. Banerjee$^{3}$, 
U. Munari$^{4}$, \newauthor
M. D. Gray$^{5,2}$,
B. Hutawarakorn Kramer$^{5,6}$,
 A.  Kraus$^{6}$ \\ 
 \\
{$^1$}Astrophysics Group, Lennard Jones Laboratory, Keele University, Keele, Staffordshire,  ST5 5BG, UK\\ 
{$^2$}Jodrell Bank Centre for Astrophysics, Department of Physics and Astronomy, University of Manchester, Manchester M13 9PL, UK\\
{$^3$}Physical Research Laboratory, Navrangpura, Ahmedabad, Gujarat 380009, India\\
{$^4$}INAF Astronomical Observatory of Padova, I-36012 Asiago (VI), Italy\\
{$^5$}National Astronomical Research Institute of Thailand, 260 Moo 4, T. Donkaew, A. Maerim, Chiangmai 50180, Thailand \\
{$^6$}Max-Planck-Institut f\"ur Radioastronomie, Auf dem H\"ugel 69, D-53121 Bonn, Germany }

\begin{document}

\date{Version of \today}

\pagerange{\pageref{firstpage}--\pageref{lastpage}} \pubyear{2020}

\maketitle

\label{firstpage}

\begin{abstract}
   There are indications that the third known eruption 
   of the recurrent nova \tcrb\ is imminent, and multi-wavelength 
   observations prior to the eruption are important to
   characterise the system before it erupts. 
   \tcrb\ is known to display the SiO fundamental vibrational
   feature at 8\mic. When the anticipated eruption occurs, it is possible 
   that the shock produced when the ejected material runs into the 
   wind of the red giant in the system may be traced using SiO maser emission.
   We have used the 100~m Effelsberg Radio Telescope to search for 
   \nucl{28}{Si}O emission in the $\upsilon=1$, $\upsilon=2$, 
   $J=1\rightarrow0$ transitions, at 43.122~GHz and 42.820~GHz 
   respectively, while the system is in quiescence.
   We find no evidence for such emission.
We set stringent $3\sigma$ upper limits of 1.66~mJy on emission in the 
$\upsilon=1, J=1\rightarrow0$ transition, and 1.72~mJy in the 
$\upsilon=2, J=1\rightarrow0$ transition, respectively, for a noise bandwith 
of 250~kHz. The corresponding limits for a 31.25~kHz bandwidth are 4.69~mJy and 4.86~mJy 
respectively. These upper limits improve on previous upper limits for this system by more
than two orders of magnitude.
\end{abstract}

   \begin{keywords}
circumstellar matter ---
     stars: individual (\tcrb) ---
    novae, cataclysmic variables
   \end{keywords}

%

\section{Introduction}

Nova eruptions arise on the surfaces of white dwarfs (WDs) in semi-detached
binary systems. The secondary star in novae, usually a late-type 
evolved star or main sequence dwarf, fills its Roche lobe and material from the 
secondary forms an accretion disc around the WD. Eventually the base of 
the accreted envelope becomes degenerate and a Thermonuclear Runaway (TNR)
ensues, resulting in a nova explosion \citep{CN2}. Up to $10^{-4}$\Msun\ 
of material, enriched in C, N, O\ldots{Ca} as a result of the TNR and the
ingestion of WD material into the burning region, is ejected explosively at several 
100s to 1000s of \vunit. Eventually mass-transfer from the secondary onto the 
WD resumes and, in time, another nova explosion occurs. All novae are ``recurrent'' 
but in some cases the eruptions in a given system recur on a {\em human}
($\ltsimeq100$~yrs) timescale; these are the ``recurrent novae'' (RNe).

Few RNe are known (see \cite{anupama08} for a list), but they can be 
sub-divided into those with short ($\ltsimeq$~a day) and those with
long ($\sim1$~yr) orbital periods. The latter generally have 
red giant (RG) secondaries, with the winds normally associated with RGs. 
RNe with RG secondaries are referred to as ``symbiotic novae''. 
When the RN in a system with a RG secondary erupts, the ejected 
material runs into, and shocks, the RG wind, and a shock 
is driven into the ejecta. This results in strong X-ray emission, 
and coronal line emission in the UV, optical and infrared (IR)
\citep*[see, e.g.,][]{evans07a,evans07b,evans07c,banerjee10,banerjee14,das,munari07},
although photoionisation is also likely to play a part in 
the production of coronal line emission \citep{munari22}.

The propagation of the shock in the case of a RN eruption in a
symbiotic nova system may provide a template for understanding
the propagation of the shock as {\em supernova} (SN) ejecta run 
into the ambient medium, a process that takes centuries as 
opposed to days--weeks for a RN: RNe provide the opportunity to
study SNe in ``fast-forward'', but with the additional twist of 
a likely density enhancement in the RN equatorial
plane that may impart a bipolar morphology to the ejecta, 

A key property of RNe is that the WD components have masses that
are close to the Chandrasekhar Limit \citep{anupama08}. 
The net accretion of material 
(i.e. accreted from the RG minus that ejected in the RN explosion)
onto the WD, if positive, may in time cause it to tip over the 
Chandrasekhar Limit and explode as a Type~Ia SN.
In furthering our understanding of RNe we may therefore get a better 
understanding of Type~Ia\,s, which are of course key in the determination
of cosmic large-scale structure \citep{perlmutter,riess}.
The study of RN eruptions therefore has applications well beyond
the confines of cataclysmic variable systems.

\section{The RN \tcrb}
    
\tcrb\ is a RN with a RG secondary \citep{anupama08};
it last erupted in 1946; its  orbital period is 227.67\,d
\citep[see, e.g.,][]{anupama08}. The recent
rise in the optical flux \citep*{munari16}, with concurrent
changes in the X-ray flux \citep{luna18}
point to the likelihood that its next eruption is 
imminent \citep{schaefer,munari16}. \cite{luna18} attributed the 
increased activity to changes in the boundary layer;
such an increase might arise from a surge in the mass-transfer 
rate from the RG to the WD. While the cause of such a surge is 
unclear, a consequence is that the amount of material in the environment
of the RG has been substantially enhanced. More recently, \cite{luna20}
have drawn attention to a striking resemblance between the pre-1946 
eruption $B$-band light curve and the current (2022) $B$ light curve. 
They suggest that the WD in \tcrb\ is currently undergoing a high 
rate of accretion, possibly due to accretion disc instabilities
that are commonly seen in dwarf novae. They predict that the next RN
eruption in \tcrb\ is likely to occur within the next 6~years.

The circumstellar envelopes (CSE) around evolved O-rich stars 
such as Mira variables and other Asymptotic Giant Branch (AGB) 
stars provide suitable conditions for maser emission 
\citep[see, e.g.,][]{gray12}, and these 
are found even in the presence of symbiotic 
interaction with a companion.  In 2009, \cite{cho} used the Yonsei 21-m 
telescope to survey 47 symbiotic stars, with on-source integration times
in the range 30--50~miutes. They detected 43~GHz SiO maser emission 
from nineteen of these, and 22~GHz H$_2$O maser emission from 
nine (seven of which also have SiO masers). The higher SiO 
maser detection rate is likely to be because the conditions for 
SiO masing occur closer to the star where the CSE is less 
likely to be disrupted by the companion 
\citep[see, e.g.,][]{gray12,richards}.  \cite{cho} did not detect 
\tcrb, with $3\sigma$ upper limits on the peak antenna temperatures 
of 60~mK in the \nucl{28}{Si}O and \nucl{29}{Si}O isotopologues, 
corresponding to $3\sigma$ upper limits on the flux density of 0.69~Jy
over a bandwidth of 31.25 kHz.  

The observations by \cite{cho} were made on 2009 November~5,
when the RG component was eclipsing the WD, so irradiation 
effects were unimportant. The signifcance of this for the 
case of \tcrb\ is that the 
RG component of the symbiotic nova V407~Cyg displays SiO maser 
emission. When V407~Cyg erupted in 2010, material
ejected in the nova explosion disrupted the maser-bearing
region around the RG, but it was re-established 
after a couple of weeks \citep{deguchi}. An alternative
interpretation is that the initial flash-ionisation 
extended to the maser region, and destroyed the conditions for 
masing; the outer wind then recombined in $\sim$~two weeks, 
and maser emission resumed. The inner
wind of the RG in V407 Cyg recombined on an $e$-folding time of 
4.0~days \citep{munari11}, and a longer interval is to be
expected in the lower density external regions from where maser 
emission originates. Whatever the cause of the destruction and 
re-establishment of the maser region, it was 
possible to use the maser to trace the progress of the shock 
through the atmosphere of the Mira component \citep{deguchi}.

The prospect of replicating the \cite{deguchi} observations 
when \tcrb\ next erupts is an exciting one. 
Such an observation, complemented by optical and IR data, would 
provide unique insight into the physics of shock evolution
in a stellar, and in particular a RN, environment. This possibility
arises from a 2007 Spitzer Space Telescope \citep{werner04,gehrz07}
InfraRed Spectrograph \citep{houck04} observation of \tcrb\
that revealed the presence of absorption in the SiO fundamental 
vibrational band at 8\mic, which had contibutions from both the 
RG photosphere and the RG wind \citep{evans19}. The SiO 
column was determined to be $2.8\times10^{17}$~cm$^{-2}$, and 
the temperature $\sim1000$~K.
It was this detection that prompted us to observe \tcrb\ 
in the $\upsilon=1,2,J=1\rightarrow0$ transitions of \nucl{28}{Si}O.

We should note, however, that a reanalysis of the Spitzer
data \citep{evans22} showed that the wind reported by
\cite{evans19} was spurious. The earlier work considered
only the contribution of the \nucl{28}{Si}O isotopologue: 
the inclusion of other SiO isotopologues 
accounted satisfactorily for the 8\mic\ absorption in terms 
of photospheric SiO only, without the need for a wind. 

Notwithstanding this later result, we report here on an
observation of \tcrb\ in the $\upsilon=1$, $\upsilon=2$, 
$J=1\rightarrow0$, transitions of \nucl{28}{Si}O.

\section{Observations}

\begin{figure*}
   \centering
   \includegraphics[width=8.5cm]{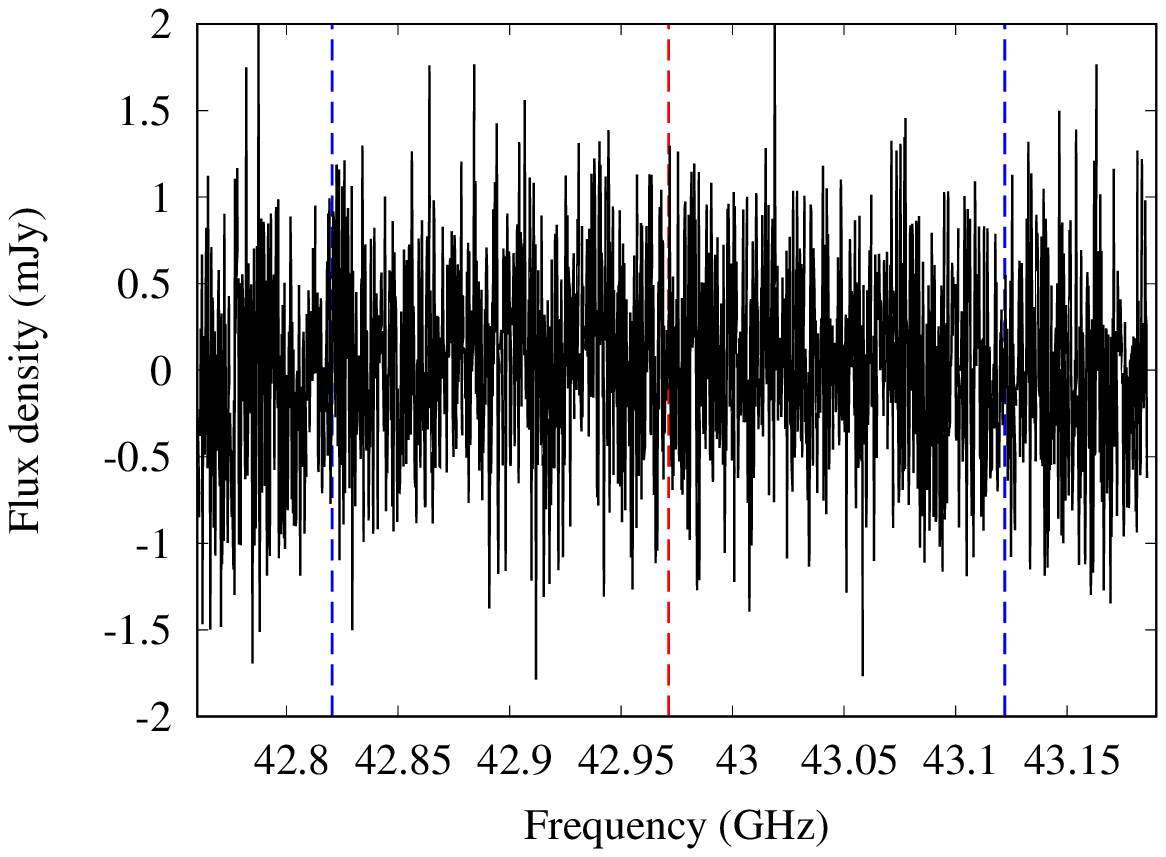}
   \put(-200,173){\color{black}{(a)}}
   \includegraphics[width=8.5cm]{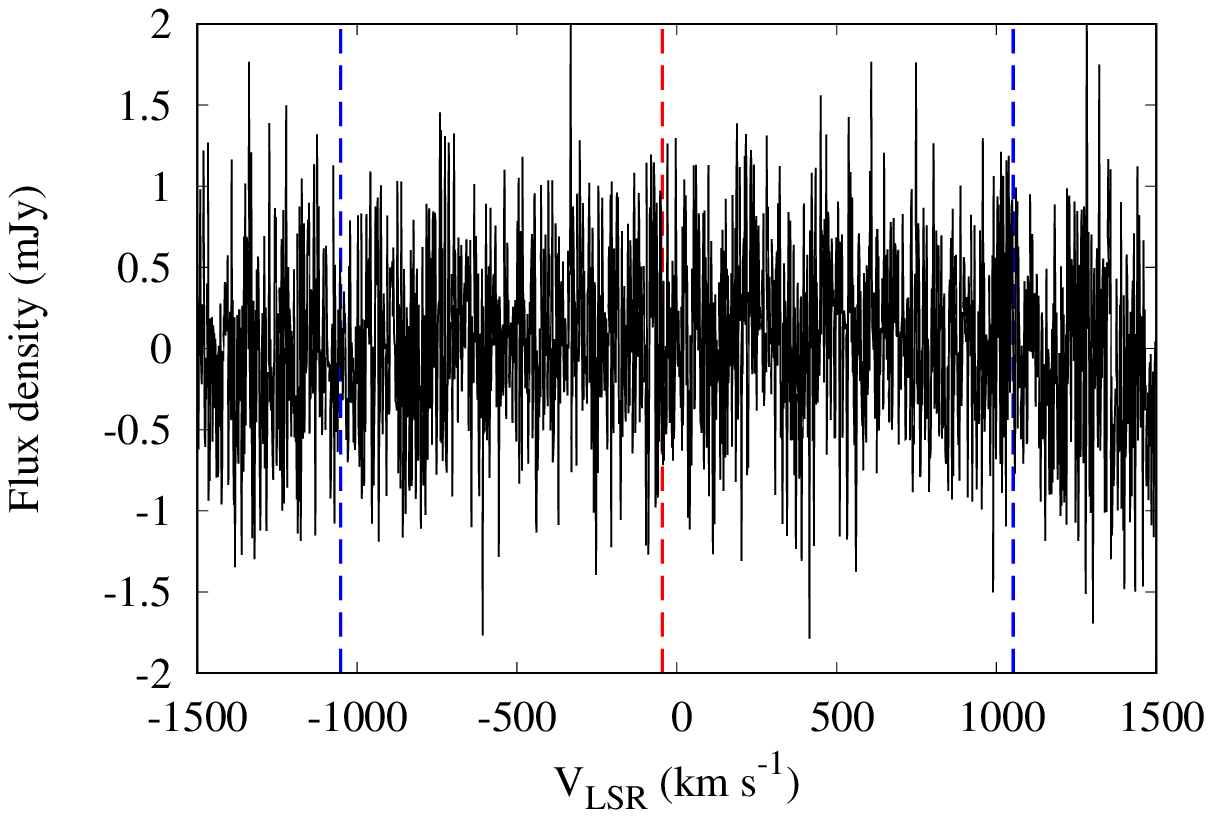}
  \put(-200,173){\color{black}{(b)}}
  \caption{(a) spectrum over the observed frequency range; data have been
   binned in 250~kHz bins. (b) as (a), but in $V_{\rm LSR}$ frame, 
   converted using the mean frequency of the two SiO transitions.
   In (b) the red vertical line denotes the $V_{\rm LSR}$ of \tcrb;
   the two blue vertical lines denote the $V_{\rm LSR}$ at which the two SiO
   transitions would be observed if present. 
   In (a), the lines correspond, in
   frequency space, to the lines in (b). \label{tcrb}}
    \end{figure*}

\tcrb\ was observed in the $\upsilon=1$, $\upsilon=2$, 
$J=1\rightarrow0$, transitions at the respective frequencies of
43.122~GHz and 42.820~GHz using the 
100~m Effelsberg Radio Telescope \citep{altenhoff80}.
The observation in the Q-band employed the 7.0\,mm secondary-focus 
receiver with the XFFS backend, with 65536 spectral channels. 
The total bandwidth was 500~MHz; the channel width in this 
configuration is 7.629~kHz (0.053\vunit\ at a frequency of 43\,GHz).
The system-equivalent flux density (SEFD) is assumed to be that measured 
at 44.1\,GHz in 2019 
December\footnote{https://eff100mwiki.mpifr-bonn.mpg.de/doku.php?id=\\information\_for\_astronomers:rx:s7mm\_db}, i.e. 140\,Jy.
The line frequency was set at 42.971335~GHz, midway between 
the frequencies of the SiO transitions. The bandwidth was such that 
both transitions would be captured, including allowances for 
uncertainties in the radial velocity.

The observations were carried out on three occasions,
2021 February 4, February 12 and February 23-24; at all these
times the WD component was eclipsed by the RG. 
This ensured that there were no complications
due to irradiation of the visible RG atmosphere by hard 
radiation from the WD that might disrupt any SiO maser zone.
However  weather conditions during the two earlier runs were poor, 
and only the data obtained on February 23-24 are presented here.
Each observation was corrected to the $V_{\rm LSR}$ appropriate 
at the time of observation, and the corrected observations
averaged. 

The heliocentric velocity of \tcrb\ is given by \cite{fekel00}
as $-27.79\pm0.13$\vunit; {\it Gaia} DR2 
\citep{gaia-a,gaia-b} gives $-27.1\pm6.9$\vunit. 
We use the \citeauthor{fekel00} value here,
to give $V_{\rm LSR} =-44.8$\vunit\ for \tcrb.
Allowance has to be made for the orbital motion of \tcrb,
although as it was observed in the above window
this would have been minimal. However, the wind velocity
\citep[$-19$\vunit\ for \tcrb;][]{munari16} ---
which will be at maximum as the RG passes at inferior conjunction ---
might also have an effect on the observed frequencies of the transitions.

\tcrb\ was observed for a total of 8~hours on February 23-24.
The same time was taken for calibration via position switching,
and the observations of standard flux calibrators.
For the flux density calibration the spectra were corrected for
the atmospheric absorption and the gain-elevation effect 
(loss of sensitivity due to the gravitational deformation of 
the main dish). The conversion factor from antenna temperature 
to flux density in Jy was determined by regular observations 
of 3C\,286.
Calibration uncertainties are estimated to be $\sim10$\%.
We note that the observing time was negligible by comparison with
the orbital period; therefore the orbital motion had
no impact on the observation. The GILDAS/CLASS2 packages
\citep{pety05} were used in the spectral line data reduction.

\section{Results}

The width of the \nucl{28}{Si}O  $\upsilon = 1, J = 1\rightarrow0$ 
transition, as measured by a single aperture, in the oxygen-rich AGB star 
R~Cas is $\sim1-2$\vunit\ \citep{assaf11}, corresponding to a 
bandwidth of 250~kHz. Our spectrum of \tcrb\ was therefore
binned into 250~kHz bins; the averaged spectrum is shown in 
Fig.~\ref{tcrb}(a), Fig.~\ref{tcrb}(b) shows the spectrum in terms of
$V_{\rm LSR}$. Clearly there is no obvious emission corresponding
to the two SiO transitions considered here. 

In order to place upper limits on the \nucl{28}{Si}O emission, 
we consider two noise bandwidths. First, we use 250\,kHz, as in Fig.~\ref{tcrb}, 
corresponding to the typical width of an SiO maser line. The 3$\sigma$ upper
limits in this case are 1.66\,mJy for the $\upsilon=1, J = 1 \rightarrow 0$ 
transition, and 1.72\,mJy in the $\upsilon=2, J = 1 \rightarrow 0$ transition. 
For a more direct comparison with the work of \cite{cho}, 
who used a 64~MHz bandwidth with 2048~channels for their SiO observations,
we use a bandwidth of 31.25\,kHz. 
Over this bandwidth our 3$\sigma$ upper limits 
are 4.69\,mJy ($\upsilon=1$) and 4.86\,mJy ($\upsilon=2$).
Given the larger aperture and the much longer integration 
times in our study, it is not surprising that our limits improve
on those reported by \cite{cho} by more than two orders of magnitude.
   
\section{Conclusions}

We have searched for 42.820~GHz and 43.122~GHz maser emission from 
\nucl{28}{Si}O in the circumstellar environment of \tcrb. We 
find no evidence for such emission. 

Nonetheless we strongly encourage a repeat of these observations
when \tcrb\ does eventually erupt. The prospect of tracing the
shock through the wind of the RG component, which 
may be confined to a ring inclined to the orbital plane
\citep*{theuns93,booth16}, is one that will surely be too good to miss.
Monitoring of \tcrb\ at these frequencies is ongoing.

\section*{Acknowledgments}
This paper is based on observations carried out with the 100-m telescope 
of the MPIfR  (Max-Planck-Institut f\"ur Radioastronomie) at Effelsberg. We are
grateful for a generous allocation of observing time.

This work has made use of data from the European Space Agency (ESA) mission
{\it Gaia} ({https://www.cosmos.esa.int/gaia}), processed by the {\it Gaia}
Data Processing and Analysis Consortium (DPAC,
{https://www.cosmos.esa.int/web/gaia/dpac/consortium}). Funding for the DPAC
has been provided by national institutions, in particular the institutions
participating in the {\it Gaia} Multilateral Agreement.

BP is supported by a Development and Promotion of Science and 
Technology Talents Project scholarship awarded by the Royal 
Thai Government.
DPKB is supported by a CSIR Emeritus Scientist grant-in-aid and 
is being hosted by the Physical Research Laboratory, Ahmedabad.


\section*{Data availability}

The data underlying this paper will be shared on a reasonable request
to the corresponding author.






\bsp	
\label{lastpage}
\end{document}